\documentclass[twocolumn,showpacs,epsf,pra]{revtex4}
\usepackage{graphicx}
\usepackage{amsmath}
\usepackage{color}
\usepackage{epsfig}
\newcommand{\ud}{\mathrm{d}}
\newcommand{\MR}{\mathcal{R}}
\newcommand{\mA}{\mathcal{A}}

\newcommand{\mr}{\mathrm}

\newcommand{\magdip}{^{\mu}}                
\newcommand{\spor}{^{\mathrm{so}}}          
\newcommand{\spsp}{^{\mathrm{ss}}}          
\newcommand{\hf}{^{\mathrm{hf}}}            
\newcommand{\hfmin}{^{\mathrm{hf}-}}        
\newcommand{\hfplus}{^{\mathrm{hf}+}}       
\newcommand{\Zeeman}{^{\mathrm{Z}}}         
\newcommand{\central}{^{\mathrm{cen}}}      
\newcommand{\exchange}{^{\mathrm{exch}}}    
\newcommand{\dispersion}{^{\mathrm{disp}}}  
\newcommand{\phiE}{^{\mathrm{E}}}           
\newcommand{\phiL}{^{\mathrm{l}}}           

\begin{document}

\hyphenation{Feshbach Fesh-bach}

\title{ Predicting scattering properties of ultracold atoms: adiabatic accumulated phase method and mass scaling }

\begin{abstract}

Ultracold atoms are increasingly used for high precision experiments that can be utilized to extract accurate scattering properties. This calls for a stronger need to improve on the accuracy of interatomic potentials, and in particular the usually rather inaccurate inner-range potentials. A boundary condition for this inner range can be conveniently given via the accumulated phase method. However, in this approach one should satisfy two conditions, which are in principle conflicting, and the validity of these approximations comes under stress when higher precision is required.
We show that a better compromise between the two is possible by allowing for an adiabatic change of the hyperfine mixing of singlet and triplet states for interatomic distances smaller than the separation radius. A mass scaling approach to relate accumulated phase parameters in a combined analysis of isotopically related atom pairs is described in detail and its accuracy is estimated, taking into account both Born-Oppenheimer and WKB breakdown. We demonstrate how numbers of singlet and triplet bound states follow from the mass scaling.
\end{abstract}

\author{B.J.~Verhaar}
\author{E.G.M.~van Kempen}
\altaffiliation[Present address: ]{Philips Applied Technologies}
\author{S.J.J.M.F.~Kokkelmans}
\affiliation{ Eindhoven University of Technology, P.O.~Box~513, 5600~MB  Eindhoven, The Netherlands }

\date{\today}


\maketitle

\section{Introduction}

In 1976 Stwalley \cite{stwalley} suggested the existence of magnetically induced Feshbach resonances in the scattering of cold hydrogen atoms. He pointed out that the specific magnetic field strengths where they occur should be avoided to achieve a stable cryogenically cooled H gas, in view of an enhanced decay at resonance. In 1992 one of the present authors (BJV) and co-workers pointed to a positive aspect of such Feshbach resonances \cite{tiesinga}: they allow for an easy control of the interaction strength between \textit{ultracold} atoms, i.e., atoms in the energy range where their interaction is limited to s-waves. In such circumstances, the interaction strength is characterized by the s-wave scattering length $a$. With a Feshbach resonance, the interactions can be tuned from weak to strong and from attractive to repulsive by simply changing an externally applied magnetic field.

Since then these resonances have become an indispensable tool in many successful attempts to control the interatomic interaction, to form ultracold molecules by associating atoms, and to create a superfluid Fermi gas. Feshbach resonances allow experiments with ultracold atoms access to a multitude of the most diverse many-body phenomena~\cite{bloch}. Systematic theoretical work to determine resonant field strengths and scattering lengths for almost all stable alkali atoms started immediately after 1992~\cite{chu,moerdijk,gardner,ritchie} and played a crucial role in the first realizations of Bose-Einstein condensation BEC in 1995~\cite{wieman,ketterle,hulet} (we will use the term alkali atom in reference to an alkali metal). An example is presented in Sec.~\ref{subsec:accumulated} in connection with the first determinations of scattering lengths. In recent years many experiments have opened the field of ultracold gases with mixed atomic species, where Feshbach resonances continue to be an indispensable tool.

A description of cold collisions between ground-state atoms (and also weakly bound states) requires highly accurate central interaction potentials. Except for the lightest elements (H and Li), \emph{ab initio} potentials do not possess the required accuracy at short range. The slightest change of a potential in that range can easily turn a positive into a negative scattering length, information which is crucial for instance to predict the stability of a BEC.

A way to account for that is to summarize the "history" of the collision for interatomic distances $r$ smaller than a separation radius $r_{0}$ by means of a boundary condition on the wave function at $r_{0}$, and to determine that condition from a restricted set of available experimental data~\cite{chu,moerdijk,gardner}. The basic philosophy of this approach is to give up the goal of extracting the detailed short-range potential as a whole from experiment in favor of a boundary condition with only a few parameters. The boundary condition takes the form of a radial phase of the zero-energy wave function accumulated in the interval $r<r_{0}$ in either the singlet or the triplet channel, and its energy and angular momentum derivatives. This presupposes pure singlet and triplet wave functions, which is justified for small interatomic distances where the singlet and triplet states are far enough apart in energy to neglect hyperfine mixing.

Over the years the accuracy of the description of scattering properties obtained with this method has shown a dramatic improvement, keeping pace with the accuracy of the measurements. In this paper we describe an extension of the accumulated phase method, the \textit{adiabatic} accumulated phase method, presented briefly in a previous publication~\cite{vankempen}, in which the hyperfine interaction is not completely neglected, but taken into account adiabatically for $r < r_0$. It enables one to describe the interaction and scattering of (ultra)cold atoms to unprecedented precision. It is also unparalleled in comprehensiveness: it allows the prediction of a large and varied set of experimental data, once the accumulated phase parameters have been determined from a restricted set of experimental data. We thus refrain from extracting the short range, deep part of the central interaction potentials, singlet and triplet, in favor of predicting new scattering properties. As such, it sets an example for future "state of the art" applications to other atoms, in particular to interactions between unlike atomic species. It should also be noted that predictive power of the accumulated phase method extends to both scattering and bound states, because the corresponding wave functions do not differ essentially at $r_0$.

As a background for the later exposition we start in Sec.~\ref{sec:IntAccum} with a summary of the accumulated phase method as we introduced and applied it previously. This section also serves to introduce our notation for the various terms in the Hamiltonian. As a last step in that section we describe our approach to include the spin-spin interaction, in particular its second-order spin-orbit part, in the accumulated phase method. The adiabatic extension of the latter method is presented in Sec.~\ref{sec:adiabaccum}. We compare the effectiveness of the new adiabatic accumulated phase method to that of the conventional method. Sec.~\ref{sec:mass scaling} deals with the subject of mass scaling. It is based on the Born-Oppenheimer and WKB approximations. In this connection it should be emphasized that the concept of mass scaling as studied here is basically different from that in other studies of cold atom scattering (see, e.g. Ref.~\cite{tiemann}), in that we apply it to a restricted range of interatomic distances thus avoiding the range $r > r_0$, in part of which the central potentials become too shallow to allow for an accurate scaling close to the dissociation energy. In Sec.~\ref{sec:accuracy mass scaling} we discuss the accuracy of the mass scaling, taking into account both Born-Oppenheimer and WKB breakdown, thus showing mass scaling to be a promising method to relate accumulated phase parameters for different isotopes of the same element. Sec.~\ref{sec:choice r0} compares the conventional and adiabatic accumulated phase methods and discusses the $r_0$ dependence of our predictions. In Sec.~\ref{sec:numbers bound states} we demonstrate our method of determining the numbers of pure singlet and triplet bound states of the various atomic species involved in the mass scaling. To our knowledge this is the only available method to extract numbers of bound states without knowing the potentials at short range. We illustrate it by means of the example of the $^{85}$Rb and $^{87}$Rb atoms. A summary and outlook are presented in Sec.~\ref{sec:summary}.

\section{Interactions and Accumulated phase method}\label{sec:IntAccum}
\subsection*{Single particle Hamiltonian}\label{subsec:single particle}
\begin{figure}
\begin{center}
\includegraphics[width=0.8\columnwidth]{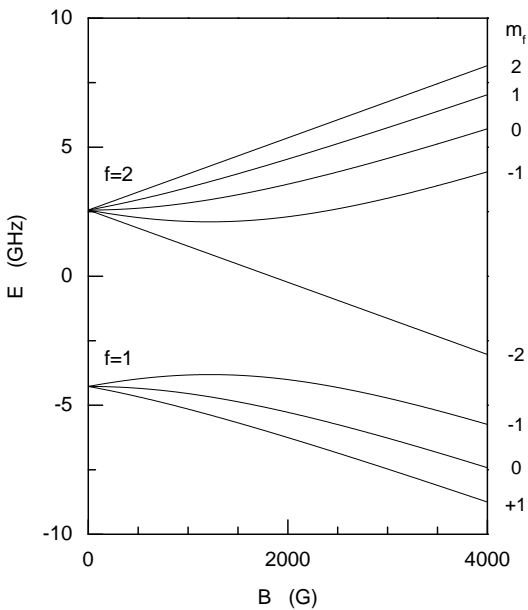}
\end{center} \caption{Hyperfine diagram of ground-state $^{87}$Rb. The energy curves of the hyperfine states are labeled by their $B$=0 quantum numbers. A similar diagram with $i = 5/2$ is valid for the $^{85}$Rb atom.}
\label{graph:hyperfine}
\end{figure}

We consider the electronic ground state of an alkali atom. The valence electron has spin $s$=$\frac{1}{2}$ while the nucleus has spin $i$, which in particular equals $\frac{3}{2}$ for $^{87}$Rb and $\frac{5}{2}$ for $^{85}$Rb. Note that lower case characters are used to indicate single atom properties while we reserve capitals for two-atom systems. In total there are thus $2(2i+1)$ possible `ground states' for an alkali atom. This degeneracy is lifted by interactions both within the atom and with external fields. To begin with, the nuclear spin interacts with the valence electron spin via the hyperfine interaction for an atom $j$:
\begin{equation}
V\hf_j=\frac{a\hf}{\hbar^2} \vec{s}_j \cdot \vec{i}_j,
\label{eq:hyperfine_splitting}
\end{equation}
with $a\hf$ the hyperfine constant (dimension of energy), leading to the hyperfine splitting according to $f = i \pm \frac{1}{2}$ ($\vec{f} = \vec{s} + \vec{i}$) with each $f$-state $(2f+1)$-fold degenerate. In turn, the remaining degeneracy is lifted when atom $j$ is placed in a magnetic field $\vec{B}$, giving rise to a spin Zeeman term
\begin{equation}
V^{Z}_j=\left( \gamma_{e,j}\vec{s}_j - \gamma_{n,j}\vec{i}_j
\right)\cdot \vec{B}, \label{eq:Zeeman_interaction}
\end{equation}
where $\gamma_{e,j}$ and $\gamma_{n,j}$ are the electronic and nuclear gyromagnetic ratios. The behavior of the valence electron in alkali atoms is influenced by the electrons filling the inner shells causing the gyromagnetic ratio $\gamma_{e,j}$ to be slightly different from that of a free electron.
Eqs.~(\ref{eq:hyperfine_splitting}) and (\ref{eq:Zeeman_interaction}) lead to the familiar graphs for the energy of the atomic hyperfine states as a function of the magnetic field $B\hat{z}$ (see Fig.~\ref{graph:hyperfine}). The single atom hyperfine states are commonly denoted by $|f,m_f\rangle$ even though $f$ is only a good quantum number for $B=0$.

\subsection*{Two particle Hamiltonian}\label{subsec:two_particle}
We consider two like alkali atoms. They experience in the first place a mutual central interaction, that can be written as
\begin{equation}\label{eq:V_central}
V\central(r)=V_S(r) P_S + V_T(r) P_T,
\end{equation}
with $P_{S,T}$ projection operators on the spin singlet and triplet subspaces and $r$ the interatomic separation. The singlet and triplet potentials differ by twice the exchange energy $V\exchange(r)$ and are at large distances given by
\begin{equation}
V_{S,T} = V\dispersion -\left(-1\right)^S~ V\exchange.
\label{eq:V_SingTrip}
\end{equation}
The dispersion energy $V\dispersion(r)$ is described by
\begin{equation}
V\dispersion = -\left(\frac{C_6}{r^6} + \frac{C_8}{r^8} +
\frac{C_{10}}{r^{10}} +  \ldots \right), \label{eq:Vdisp}
\end{equation}
with the dispersion coefficients $C_n$. An analytic expression for the exchange energy in Eq.~(\ref{eq:V_SingTrip}) has been derived by Smirnov and Chibisov \cite{smirnov} for $r$ values where the overlap of the electron clouds is sufficiently small:
\begin{equation}
V\exchange = \frac{1}{2}J r^{\frac{7}{2\kappa}-1} e^{-2\kappa r}.
\label{eq:Vexch}
\end{equation}
In this equation $J$ and $\kappa$ are positive constants with $\kappa^2/2$ the atomic ionization energy; $r$, $J$, and $\kappa=0.554....$ in atomic units. The most recent value for $J$ was given by Hadinger \emph{et al}.~\cite{hadinger}, who made use of Ref.~\cite{radzig}.

Leaving out the center of mass kinetic energy and including the above interactions the total effective Hamiltonian for two colliding ground-state alkali atoms becomes
\begin{equation}\label{eq:total_hamiltonian}
H=\frac{\vec{p}~^2}{2\mu}+\sum^2_{j=1} \left( V\hf_j+V\Zeeman_j
\right) + V\central,
\end{equation}
in which the first term represents the kinetic energy with $\mu$ the reduced mass and $\vec{p}$ the relative momentum operator.

The hyperfine term can be written as the sum of two parts with different symmetry with respect to interchange of the electronic or nuclear spins,
\begin{eqnarray}\label{eq:hf_parts}
V\hf&=&\frac{a\hf}{2\hbar^2}(\vec{s}_1+\vec{s}_2)\cdot(\vec{i}_1+\vec{i}_2)
+
\frac{a\hf}{2\hbar^2}(\vec{s}_1-\vec{s}_2)\cdot(\vec{i}_1-\vec{i}_2)\nonumber \\
&\equiv& V\hfplus+V\hfmin.
\end{eqnarray}
The convenience of this splitting arises from the fact that $V\hfplus$ is diagonal in $S$, whereas $V\hfmin$, being antisymmetric in $\vec{s}_1$ and $\vec{s}_2$, is the part coupling singlet and triplet states.

For the interactions mentioned up to now the total Hamiltonian $H$ is invariant under independent rotations of the spin system and the orbital system around the axis through the overall center of mass parallel to the magnetic field. Therefore the projection of the total spin angular momentum $\vec{f_1} + \vec{f_2} \equiv \vec{F}$ and the orbital angular momentum $\vec{l}$ along this axis are separately conserved during the collision. Since $V\central$ depends on $r$ only and not on $\hat{r}=\vec{r}/r$, $\vec{l}$ is even conserved as a 3D vector. As a consequence, $m_F$ and the rotational quantum numbers $l$ and $m_l$ are good quantum numbers.

Two other interactions influence the two-atom system, which are much weaker than the above-mentioned effects, but nevertheless can play a significant role in interpreting cold atom experiments. The first one is a direct interaction between the spins of the valence electrons via their magnetic dipole moment. It is given by
\begin{equation}\label{eq:magnetic_spin_spin_interaction}
V\magdip (\vec{r}) = \mu_0 \frac{\vec{\mu}_1 \cdot \vec{\mu}_2 -
3(\vec{\mu}_1 \cdot \hat{r})(\vec{\mu}_2 \cdot \hat{r})} {4\pi r^3},
\end{equation}
with $\mu_0=4\pi\cdot 10^{-7}$Hm$^{-1}$ and $\vec{\mu}_j$ the electronic magnetic dipole moment of atom $j$. We leave out the much weaker magnetic dipole interactions in which the nuclear magnetic moments are involved. Another interaction with similar structure arises as a second order effect of the spin-orbit interactions of the valence electrons~\cite{mizushima} via an intermediate coupling to electronically excited states. The well-known electronic (first-order) spin-orbit interactions do not play a role for the orbital s-states of the valence electrons.

In total, we thus have a spin-spin interaction $V\spsp$ between the valence electrons, consisting of two parts:
\begin{equation}\label{eq:Vss_sum}
V\spsp = \left(V\spsp)\magdip +(V\spsp\right)\spor,
\end{equation}
a magnetic dipole part and a part arising from the spin-orbit interaction in second order. The dipole part, when expressed in the spin vectors $\vec{s}_i$ is given by
\begin{equation}\label{eq:Vss_tensorproduct}
\left(V\spsp\right)^{\mu} = \frac{\mu_0 \gamma^2_e}{4 \pi r^3}%
\left[\vec{s}_1 \cdot \vec{s}_2 - 3 (\vec{s}_1 \cdot \hat{r})%
(\vec{s}_2 \cdot \hat{r}) \right].
\end{equation}
The part $(V\spsp)\spor$ has effectively the same spin-angle structure (the factor between square brackets), but is multiplied by a different radial factor. This factor has been calculated via an \textit{ab initio} electronic structure calculation by Mies \emph{et al}.~\cite{mies} and can be approximated as an exponentially decaying effective form $f(r)$ for increasing $r$:
\begin{equation}
f(r)=-\frac{1}{\hbar^2}C^{so}{\alpha}^{2}e^{-\beta(r-r_{so})},  \label{eq:vso}
\end{equation}
with $\alpha$ the fine structure constant and $C^{so}$, $\beta $ and $r_{so}$ fit parameters to the \emph{ab initio }results. In the following we will show that its effect on two-atom bound or scattering states can effectively be accounted for via one parameter only~\cite{kokkelmans1,kokkelmans}, which has the form of an integral of $f(r)$. The total $V\spsp$ apparently has the spin-angle structure of a scalar product of two irreducible spherical tensors of rank 2~\cite{messiah} (Sec.~31):
\begin{eqnarray}
\left( \vec{s}_{1},\vec{s}_{2}\right) _{2}\cdot \left( \hat{r},\hat{r}%
\right) _{2} &=&\sum_{\sigma =-2}^{2}(-1)^{\sigma }\left( \vec{s}_{1},\vec{s}%
_{2}\right) _{2\sigma }\left( \hat{r},\hat{r}\right) _{2-\sigma }
\label{eq:scalar product} \nonumber \\
&\sim &\sum_{\sigma =-2}^{2}(-1)^{\sigma }\left( \vec{s}_{1},\vec{s}%
_{2}\right) _{2\sigma }Y_{2-\sigma }(\hat{r}).
\end{eqnarray}
As a consequence, it is invariant under the simultaneous 3D rotations of the orbital and spin degrees of freedom, thus conserving the total two-atom angular momentum. On the other hand, it is not invariant under independent orbital and spin rotations. It therefore obeys triangle type $S$ and $l$ selection rules for a second rank tensor: it couples only spin triplet states and it couples for instance the $l$ = 0 and 2 relative partial waves of the two atom system.

As mentioned before, both parts of the spin-spin interaction are relatively weak. In most applications they can safely be neglected. That these interactions cannot always be neglected is illustrated by the observation of Feshbach resonances in $^{133}$Cs~\cite{chu1} and $^{87}$Rb~\cite{marte} experiments in which colliding ultracold atoms, approaching each other in an $s$-wave, resonate with an $l$=2 or even $l$=4 (quasi-)bound state coupled via the spin-spin interaction to the entrance channel (in the $l$=4 case this interaction is needed twice: $s$-wave $\leftrightarrow$ $d$-wave $\leftrightarrow$ $g$-wave).

\subsection*{Accumulated phase method}\label{subsec:accumulated}

\begin{figure}
\begin{center}
\includegraphics[width=0.45\textwidth]{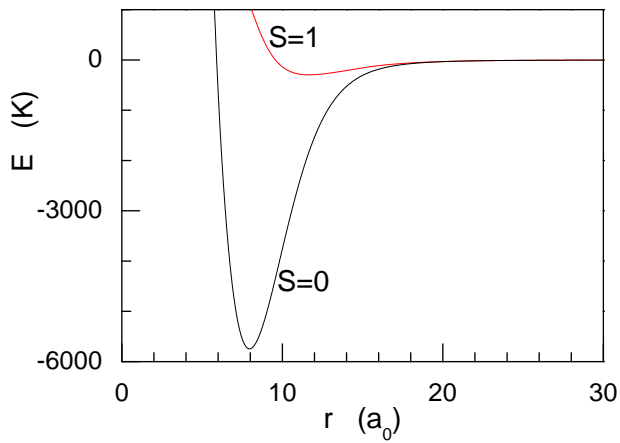}
\end{center} \caption{(Color online) Singlet ($S=0$) and triplet ($S$=1) potentials for a pair of  rubidium atoms in the electronic ground state.} \label{graph:VsingVtrip}
\end{figure}

\begin{figure}
\begin{center}
\includegraphics[width=0.45\textwidth]{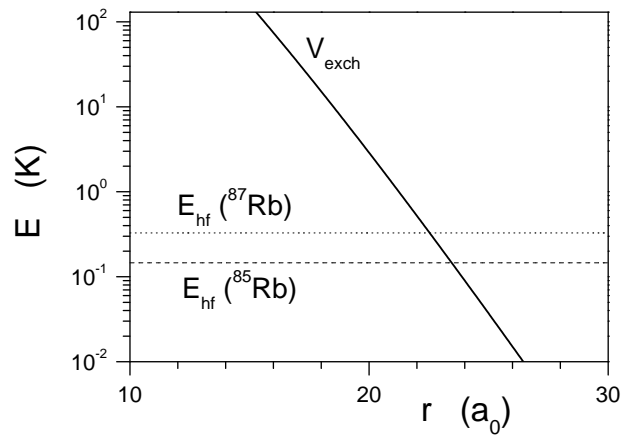}
\end{center}
\caption{ The $S=0$ $\leftrightarrow$ $S=1$ energy splitting of two ground-state rubidium atoms (equal to $2 V\exchange$) versus interatomic separation. The hyperfine energies for the isotopes $^{85}$Rb and $^{87}$Rb are indicated for comparison.} \label{graph:VExchVhf}
\end{figure}

As a background for the later exposition of its new adiabatic variant and the mass scaling procedure, it is useful first to recapitulate the basic features of the conventional accumulated phase method. The separation radius $r_0$ where we impose the boundary condition is chosen so as to fulfill three conditions:
\begin{enumerate}
\item $r_0$ should be so small that in the range $r < r_0$ the lowest $S=0$ and $S = 1$ two-atom electron states (see  Fig.~\ref{graph:VsingVtrip} and Fig.~\ref{graph:VExchVhf} for a pair of Rb atoms) are sufficiently far apart in energy for the singlet-triplet coupling due to $V\hfmin$ to be negligible. This makes it possible to formulate the boundary condition in terms of pure singlet and triplet waves.
\item On the other hand $r_0$ has to be so large that the singlet and triplet potentials for atomic distances $r>r_0$ can be accurately described by their asymptotic form $V\dispersion \mp V\exchange$ according to Eqs.~(\ref{eq:Vdisp}) and (\ref{eq:Vexch}), with a small number of unknown parameters.
\item The value of $r_0$, as well as both the energy $E$ relative to threshold and the angular momentum $l$ values playing a significant role in the experimental data should be small enough that a rapidly converging expansion of the $S=0$ and $S=1$ phases in powers of $E$ and $l(l+1)$ is possible, thus also containing a small number of unknown parameters.
\end{enumerate}

In view of the possibility that these conditions are contradictory, it is far from obvious that a suitable $r_0$ value can be found. In the first half of the nineties when three U.S. experimental groups attempted to create a BEC in an alkali atomic gas, it was possible to predict the signs and (in some cases rough) magnitudes of the scattering lengths for almost all alkali species, determining the stability ($a > 0$) or instability ($a < 0$) of a large BEC. This essential information could be obtained with the accumulated phase method using a value 19 and even 20$a_0$ for $r_0$ ($a_0$ = Bohr radius = 0.5291772 $\times 10^{-10}$m). For example, a predicted negative $a$ for $^{85}$Rb and a positive $a$ for $^{87}$Rb atoms~\cite{gardner} (both spin-stretched) led Wieman, Cornell and co-workers in 1995 to switch from $^{85}$Rb to $^{87}$Rb in their experiment, leading to the first successful realization of BEC in an ultracold atomic gas~\cite{wieman}.

\begin{figure}
\begin{center}
\includegraphics [width=0.45\textwidth]{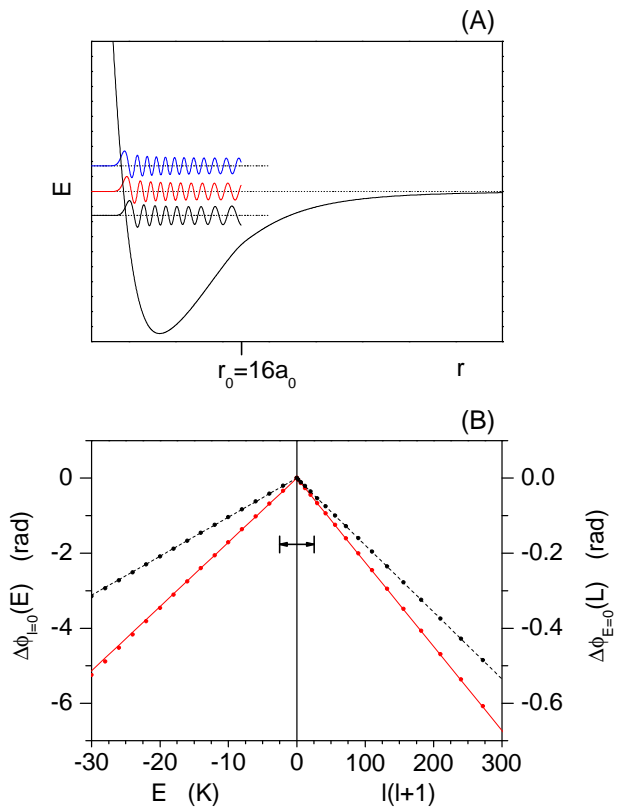}
\end{center} \caption{(Color online) Part A illustrates the behavior of the wave function's phase near $r_0 = 16 a_0$ for three different energies. A comparison between the true accumulated phase (dots) and a first order approximation (solid lines for triplet, dashed lines for singlet) is shown in part B. As a function of $E$ and $l(l+1)$ the graph shows the difference in accumulated phase $\phi(E,l)$ at $r=r_0$ as compared to the $E=0$, $l=0$ situation: $\Delta\phi(E,0)=\phi(E,0)-\phi(0,0)$ and $\Delta\phi(0,l)=\phi(0,l)-\phi(0,0)$, respectively. The horizontal arrow indicates the typical $E$ and $l$ ranges for which we apply the first order approximation. Typical rubidium potentials are used for this calculation. Note that for clarity the energy intervals for the wave functions in part A exceed the energies occurring in practice by far.}
\label{graph:GeaccPhase}
\end{figure}

The concept of an accumulated phase was originally introduced in the spirit of the WKB approximation as the local phase of a rapidly oscillating radial wave function at $r_{0}$. Its value $\phi_S(E,l)$ and $\phi_T(E,l)$ for each of the singlet and triplet wave functions is defined by
\begin{equation}
\psi(r_0) =A \frac{\sin [ \phi(E,l)]}{\sqrt{k(r_0)}}
\label{eq:wavefunction}
\end{equation}
and its radial derivative, with up to a constant the singlet or triplet accumulated phase
\begin{equation}
\phi(E,l) = \int^{r_0}k\left(r\right) dr.
\label{eq:phaseintegral}
\end{equation}
Here $k\left( r\right) $ is the local radial wave number for the channel involved:
\begin{equation}
k^{2}\left( r\right) =\frac{2\mu }{\hbar ^{2}}\left[ E-V\left(
r\right) -\frac{\hbar ^{2}l\left( l+1\right) }{2\mu r^{2}}\right] \,
\label{eq:k2}
\end{equation}
with $\mu$ the reduced mass and $V(r)$ the singlet or triplet potential.
With respect to condition (3) earlier in this section we repeat that for (ultra)cold colliding atoms ($T \lesssim 1\mu$K) and near-dissociation bound states we are most often considering, $E$ is close to 0 (compared to the depth of the potential at $r_0$) and $l$ is at most 4. As shown in Fig.~\ref{graph:GeaccPhase} for Rb atoms, the small $E$ and $l$ ranges then allow a first order Taylor expansion for $\phi(E,l)$ according to:
\begin{eqnarray}
\phi(E,l)&=&\phi(0,0)+\frac{\partial \phi}{\partial E}E+%
\frac{\partial \phi}{\partial \lbrack l(l+1)]}l(l+1) \nonumber \\
&\equiv& \phi^0 + E\phi^E + l(l+1)\phi^l.
\label{eq:phase expansion}
\end{eqnarray}
The generally fractional s-wave vibrational quantum numbers at dissociation, $v_{DS}$ and $v_{DT}$, are essentially equivalent to the zero-order Taylor terms.
They provide for more direct physical insight, however, being a measure of how close the last bound or the first unbound two-atom state is to the dissociation threshold. Their fractional values are defined via interpolation between successive infinite values of the scattering length making use of the radial phase in the deepest part of the potential \cite{chu}:
\begin{equation}
v_D({\rm mod~1})=\frac{\phi^0-\phi^0(a=\infty)}{\pi},
\end{equation}
where $\phi^0(a=\infty)$ would be consistent with an infinite value of the scattering length, i.e. a potential which has a bound state at the dissociation threshold. The energy-derivatives correspond to the classical sojourn time
\begin{equation}\label{eq:tau}
\tau_{col} = 2\hbar \partial \phi/\partial E
\end{equation}
of the atoms in the distance range $r < r_{0}$ for $l = 0$ and energies close to threshold. The $l(l+1)$ derivatives are a measure for the influence of the centrifugal force in the rotating two-atom system.

It is very convenient and intuitively appealing to define the boundary condition in the above way. The validity of the WKB approximation, however, is not strictly necessary, since the phase $\phi(E,l)$ can be defined in terms of a logarithmic derivative. For $r>r_{0}$ there is a coupling region where the exchange interaction is of similar magnitude as the hyperfine and Zeeman energies, as indicated in Figs.~\ref{graph:VsingVtrip} and \ref{graph:VExchVhf} for the Rb atoms. For larger interatomic distances where $V^{exch}$ has further decreased the two-particle hyperfine states form a good basis.

An advantage of the accumulated phase method compared to alternatives~\cite{tiesinga1,julienne,gao} is that the above set of phase parameters can be systematically extended by taking more terms in the expansion (\ref{eq:phase expansion}) into account. We also point to the difference with Multichannel Quantum Defect Theory (MQDT) methods in general: in our case the scattering channels are still coupled by the exchange interaction in part of the exterior region $r > r_0$ where $V\exchange$ is of similar magnitude as the hyperfine energy, as indicated in Fig.~\ref{graph:VsingVtrip} and \ref{graph:VExchVhf}.

\subsection*{Inclusion of second-order spin-orbit interaction}\label{subsec:vso2}

In the above discussion we have neglected the spin-spin interactions $(V\spsp)\magdip$ and $(V\spsp)\spor$, which are responsible for corrections to the scattering behavior and lead to decay. The magnetic dipole interaction $(V\spsp)\magdip$ contributes mainly due to its $1/r^{3}$ long range behavior. In the range $r < r_{0}$ the indirect spin-spin interaction $(V\spsp)\spor$ can be very strong, exceeding the dipole interaction by several orders of magnitude, but it is rather weak beyond $r_{0}$. The influence of $(V\spsp)\spor$ can be taken into account in a model-independent way via one additional parameter. This can be seen as follows. In the first place we note that we can account for the $(V\spsp)\spor$ mixing of the triplet channels by means of a local $S$-matrix~\cite{thijssen}, $\underline{\underline{S}}(r)$, that will be part of the boundary condition and specifies the ratio of the outgoing and incoming parts of the total wave function at $r$. In the vicinity of $r = r_{0}$, classically accessible so that the local channel wave numbers are real and positive, the radial solutions without spin-spin interaction are given by
\begin{equation}
F_{i}(r)=\frac{\sin \left( \int_{r_{0}}^{r}k\left( r\right) \,dr+\phi
_{i}\right) }{\sqrt{k_{i}\left( r\right) }},
\end{equation}
with

\begin{equation}
 \phi _{i}=\phi
_{S/T}\left( E_{tot}-\varepsilon _{i},l_{i}\right),
\end{equation}
where the channels $i$ differ from each other by their singlet/triplet character, their $l$ values or their internal energies $\varepsilon _{i}$. To formulate a local $S$-matrix at $r_{0}$ we introduce a complementary solution
\begin{equation}
G_{i}(r)=-\frac{\cos \left( \int_{r_{0}}^{r}k\left( r\right) \,dr+\phi
_{i}\right) }{\sqrt{k_{i}\left( r\right) }},
\end{equation}
satisfying the Wronskian condition $W[F_{i},G_{i}]\equiv F_{i}G_{i}^{\prime }-F_{i}^{\prime}G_{i}=1$. We calculate the influence of $(V\spsp)\spor$ by means of a solution of the coupled Schr\"{o}dinger equation for $r < r_{0}$ using Eq.~(\ref{eq:vso}) for the form factor or any alternative. Near $r_{0}$ we expand that solution in $F$ and $G$ functions according to:
\begin{equation}\label{eq:cmatrix}
\underline{\underline{u}}(r)=\underline{\underline{F}}(r)+\underline{%
\underline{G}}(r)\underline{\underline{C}}\quad (r\approx r_{0}),
\end{equation}
with the non-vanishing elements of the diagonal $\underline{\underline{F}}$- and $\underline{\underline{G}}$-matrices corresponding to the $F$ and $G$ functions. This defines a local generally non-diagonal $\underline{\underline{C}}$-matrix that accounts for the $(V\spsp)\spor$ interaction. It is independent of the accumulated phases and depends only on the envelope of the rapidly oscillating $F$ functions. Note that we could have used complex ingoing and outgoing exponentials as basis functions instead of cosine and sine functions. The resulting complex $\underline{\underline{S}}(r=r_0)$-matrix has a simple relation with $\underline{\underline{C}}$.

\begin{figure}
\begin{center}
\includegraphics[width=0.5\textwidth]{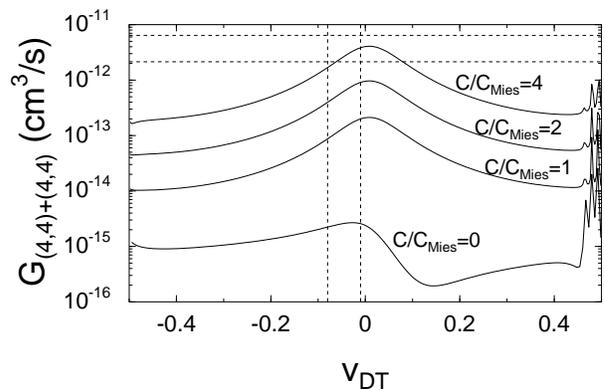}
\end{center}
\caption{Rate coefficient $G$ for the decay of Cs atoms in the $|4,4\rangle$ hyperfine state, as a function of the potential parameter $v_{DT}$ for several strengths of $(V\spsp)\spor$. The experimental boundaries for $G$ are indicated by two horizontal dashed lines, the theoretical range for $v_{DT}$ ~\cite{kokkelmans1,kokkelmans} by two vertical dashed lines.}
\label{graph:vsort44}
\end{figure}

It turns out that this $C$-matrix can to very good approximation also be obtained without solving a coupled channels equation in the range $r<r_{0}$:\ a constant times the matrix representation of the spin-angular part (\ref{eq:scalar product}), independent of the radial form factor $f(r)$, gives the same result. This can be understood as follows. As calculations show, $\underline{\underline{C}}$ depends on the spin-spin interaction only in first order. A perturbation argument leads to the conclusion that the elements $C_{ij}$ of $\underline{\underline{C}}$ have the form of a radial integral of $f(r)$ multiplied by a product $F_{i}(r)F_{j}(r)$~\cite{thijssen}. Since the (triplet) $F$-functions are almost equal over the range where $(V\spsp)\spor$ contributes significantly, the matrix representation of the spin-angular part (\ref{eq:scalar product}) is the only part in which the dependence on $i,j$ survives. We conclude that we can completely leave the detailed radial form of $(V\spsp)\spor$ out of consideration and specify only a single multiplicative constant $C$, its effective strength, to be compared to  $C_{Mies}$, the strength following from the \emph{ab initio} calculation by Mies \emph{et al.}~\cite{mies}.

An application of this approach to Cs atoms is shown in Fig.~\ref{graph:vsort44}. For atoms lighter than Rb, the magnetic dipole-dipole interaction $(V\spsp)\magdip$  adequately describes the decay of ultracold spin-stretched alkali gases. The second-order spin-orbit interaction $(V\spsp)\spor$, however, increases even more strongly with $Z$ than the first-order spin-orbit coupling and is expected to dominate the influence of $(V\spsp)\magdip$ for cesium. The figure shows the experimental rate coefficient $G$ for the decay of the spin-stretched state ($|f,m_{f}=4,4\rangle $), measured by S\"oding \emph{et al.}~\cite{soeding}. It also shows our calculated rate coefficient as a function of $v_{DT}$ for several values of $C/C_{Mies}$. The theoretical range for $v_{DT}$ is taken from \cite{kokkelmans1}. Clearly, the magnetic dipole interaction on its own ($C = 0$) is far too weak to account for the measured decay rate. The strength constant $C$ of $(V\spsp)\spor$ has to be a factor of about 4 larger than the \emph{ab initio} value $C_{Mies}$ to obtain agreement with the experimental rates~\cite{kokkelmans,remark}.

\section{Adiabatic accumulated phase method}\label{sec:adiabaccum}

The theoretical precision needed for the ``state of the art'' BEC and Fermi degeneracy experiments forces us to shift $r_0$ to atomic distances significantly smaller than $16a_0$ to neglect the singlet-triplet coupling for $r < r_0$ according to the above-mentioned condition (1) for the applicability of the straightforward accumulated phase method. We then run a real risk of violating condition (2), however. In this section we present a more sophisticated variant of the accumulated phase method, already introduced briefly in Ref.~\cite{vankempen}, that allows us to relax condition (1) to some extent, making it possible to find a value for $r_0$ while achieving the desired accuracy.

\begin{figure}
\begin{center}
\includegraphics[width=0.45\textwidth]{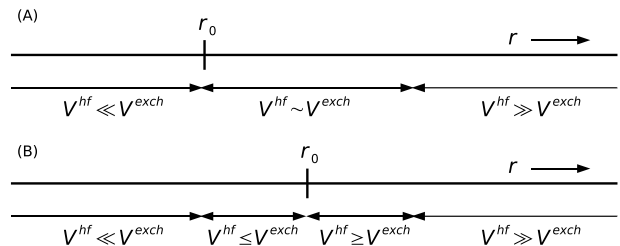}
\end{center}
\caption{Subdivision of radial ranges to illustrate choices of $r_0$. Part A distinguishes three radial ranges. In the left interval $S$ is a good quantum number. In the right interval the individual atomic hyperfine labels $f_1, m_{f1}, f_2, m_{f2}$ characterize the spin states. Conventionally, $r_0$ is chosen as far right as possible in the $V\hf \ll V\exchange$ interval. Part B shows the radial intervals as they occur in the \textit{adiabatic} accumulated phase method. The intermediate radial interval is subdivided in one in which the influence of $V\hfmin$ is small and adiabatic and one in which it is not. The separation radius $r_0$ is chosen as far right as possible in the former interval.} \label{graph:RadialRanges}
\end{figure}

In Fig.~\ref{graph:RadialRanges} we explain the difference between the conventional accumulated phase method and the new approach, distinguishing several intervals along the $r$ axis according to the relative magnitudes of $V\hf$ and $V\exchange$. In part A we consider three intervals illustrating the conventional method. In the left interval $V\hf$ is so weak compared to $V\exchange$, i.e., to the $S=0$ $\leftrightarrow$ $S=1$ splitting of potential curves, that the coupling due to $V\hfmin$ can be neglected. We thus have $S$ = 0 and 1 as a good quantum number. The remaining part $V\hfplus$, together with the two-atom Zeeman interaction $V\Zeeman$, can therefore be included effectively in the Hamiltonian via its eigenvalues, that can simply be added to the singlet and triplet potentials, in addition to their centrifugal $l$ splitting. We thus have a set of singlet and a set of triplet potential curves, each with known energy separations independent of $r$. In the right interval of part A the situation with respect to the relative magnitude of $V\hf$ and $V\exchange$ is opposite and the individual atomic hyperfine labels $f_1, m_{f1}, f_2, m_{f2}$ characterize the spin states. In the middle interval the two potential terms are comparable. The separation radius $r_0$ is chosen as far right as possible in the $V\hf \ll V\exchange$ interval. The boundary conditions for the pure singlet and triplet radial wave functions at $r_0$ along the potential curves can therefore be formulated simply in terms of $E$ and $l$ dependent pure singlet and triplet phases $\phi(E,l)$ (note that $E$ is the energy relative to dissociation).

The new insight leading to our alternative approach concerns the role of $V\hfmin$. Let us turn to part B of Fig.~\ref{graph:RadialRanges} and consider what happens when we move into the region where $V\hf \sim V\exchange$. One will first pass through an interval where the $V\hfmin$ coupling is not negligible but still small and adiabatic. In principle, $V\hfmin$ induces both a spin mixing between the $S$ = 0 and 1 states, and a perturbation on the radial wave functions. We include the spin mixing, but neglect the radial perturbation, so that the radial functions are still decoupled singlet and triplet waves characterized by pure singlet and triplet accumulated phases. Note that the spin mixing is a first order perturbation, whereas the energy perturbation on the singlet and triplet states  is a second order effect.

As a further illustration of the difference between the two methods we discuss the example of $^{87}$Rb + $^{87}$Rb scattering with initial spin state $| f_1, m_{f1}, f_2, m_{f2} >$ = $|1, -1, 1, -1 >$. We calculate the matrix $\underline{\underline{C}}$ of Eq.~(\ref{eq:cmatrix}), but instead of taking it to arise from the spin-spin interactions in the range $r < r_0$, we evaluate how it builds up from $V\hfmin$ by solving a coupled channels equation based on $V\central + V\hf + V\Zeeman$ for this radial regime. We note that in the present application of Eq.~(\ref{eq:cmatrix}) the $F$ and $G$ waves continue to arise from the part $V\central + V\hfplus + V\Zeeman$ of the Hamiltonian.

\begin{figure}
\begin{center}
\includegraphics[width=0.45\textwidth]{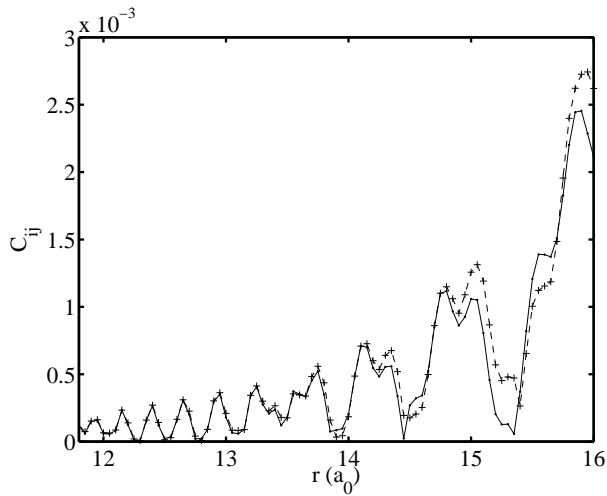}
\end{center} \caption{Comparison of conventional accumulated phase method and new alternative approach for $^{87}$Rb + $^{87}$Rb scattering with initial spin state $| f_1, m_{f1}, f_2, m_{f2} >$ = $|1, -1, 1, -1 >$. Solid line: largest $C$ matrix element $C_{ij}$ in absolute value for full coupled channel calculation in $r_0$ range [11.65, 16.0] $a_0$. Dashed line: analogous result from adiabatic accumulated phase method. Conventional method corresponds to $C_{ij}=0$}
\label{graph:cmatrix}
\end{figure}

The solid line in Fig.~\ref{graph:cmatrix} shows the largest $C$ matrix element in absolute value for $r_0$ values in the range [11.65, 16.0] $a_0$ from the latter calculation. The dashed line is the analogous quantity following from the adiabatic accumulated phase method. Clearly, the latter is in excellent agreement with the rigorous result for the small $r_0$ values. The error gradually grows to 0.25 $\times$ 10$^{-3}$ at $r_0$ = 16.0 $a_0$. This amounts to an error of about 10 \%  of the total effect due to $V\hfmin$, which by itself is of order 0.4 \% of the analogous $V\hfplus$ quantity $\phi^E \times E^{hf}(^{87}Rb)\sim$ 0.6. Note that the conventional accumulated phase method corresponds to $C_{ij}=0$.

We emphasize that the new approach includes effectively the adiabatic spin mixing in the complete range $r<r_0$. Although we impose the boundary condition that starts the coupled channel calculation in the range $r>r_0$ only at $r_0$, by its local character the adiabatic spin mixing may be understood to have been included for all smaller $r$ values. This is clearly illustrated in Sec.~\ref{sec:choice r0}, where we discuss an application of the adiabatic accumulated phase method to $^{85}$Rb and $^{87}$Rb, previously presented in Ref.~\cite{vankempen}. It turns out (see column C of table \ref{table1} in the following) that the deduced potential parameters and $v_{DS}, v_{DT}$ are highly independent of $r_0$ over a rather long range. An important aspect is a comparison with the straightforward accumulated phase method. In particular, we will present convincing evidence, in addition to Fig.~\ref{graph:cmatrix}, that the new variant allows us to shift $r_0$ to larger interatomic distances without significant loss of accuracy, thus enabling us to use more reliable potential terms in the range of interatomic distances $r > r_0$ in the form of dispersion and exchange expressions with a small number of parameters.

\section{Mass scaling: Explicit isotopic dependence of phase parameters}\label{sec:mass scaling}

As long as experimental data are analyzed for bound states and cold collisions of a single pair of (un)like atoms, it is only the local phase at $r_0$, i.e., the modulo $\pi$ part of the accumulated phase $\phi(E,l)$ that plays a role in the radial boundary condition. In this section we consider the combined analysis of several isotopic versions of atom pairs and the advantages of mass scaling in that connection. We believe that this subject will play an increasingly important role in cold atom physics, also for collisions of unlike atoms \cite{inguscio}. When analyzing  experimental data for two isotopic pairs, making use of the first terms of the Taylor expansion (\ref{eq:phase expansion}), we would need to introduce a set of 2 ($S=0,1$) times 3 ($\phi^0$, $\phi\phiE$, and $\phi\phiL$) independent parameters for each of the two-atom systems, to be determined by comparing theoretically predicted to experimentally determined properties of cold collisions or weakly bound states.

The mass scaling is based on both the Born-Oppenheimer and WKB approximations. The former approximation enables us to assume equal central potentials for the isotopic pairs. Clearly, it is essential for this approach that Born-Oppenheimer breakdown corrections can be neglected. The WKB approximation makes it possible to use an explicit expression for the accumulated phases as radial integrals containing the reduced mass via the wave number $k(r)$. As we will see, the actual value for $r_0$ chosen in applications of the adiabatic accumulated phase method is at small enough interatomic distances along the outer slope of the potentials wells for the relative atomic motion to provide for an accurate validity of the WKB approximation in the radial range $r < r_0$. We start from the WKB integral (\ref{eq:phaseintegral}) above, written more specifically as
\begin{equation}
\phi(E,l) = \int_{r_t}^{r_0}k\left(r\right) dr + \frac{\pi}{4},
\label{eq:phaseintegral1}
\end{equation}
with $r_t$ the inner turning point and the added constant $\pi/4$, associated with the quantum mechanical penetration into the inner wall of the potential~\cite{messiah}(Ch.~VI). We thus have the proportionalities
\begin{equation}
\phi^0 - \frac{\pi}{4} \propto \sqrt{\mu},
\label{eq:WKB Phi0 formula}
\end{equation}
and by differentiation of Eq.~(\ref{eq:phaseintegral1}) with respect to $E$ and $l(l+1)$:
\begin{equation}
\phi\phiE \equiv \left. \frac{\partial \phi}{\partial E}
\right|_{l=0} = \int \frac{\mu \ud r} {\hbar^2 k} \propto \sqrt{\mu},
\label{eq:phaseintegral2}
\end{equation}
\begin{equation}
\phi\phiL \equiv \left. \frac{\partial \phi}{\partial l(l+1)}
\right|_{E=0} = \int \frac{\ud r} {2 k r^2} \propto \frac {1}
{\sqrt{\mu}}.
\label{eq:phaseintegral3}
\end{equation}

Clearly, the advantages of a combined analysis of isotopes and the associated mass scaling are
a) We extend the set of available experimental data without increasing the number of fit parameters: we need the phase parameters of only one of the isotope pairs.
b) Via the scaling of $\phi^0$ the fit becomes sensitive to the number of nodes of the radial wave function left of $r_0$, in addition to the modulo $\pi$ part of $\phi^0$. With the dispersion + exchange parameters deduced in the analysis we then also know the number of nodes on the right-hand side and thus the numbers of bound singlet and triplet states for all possible isotope pairs, not only those analyzed. We will see an example of this approach in the case of $^{85}$Rb+$^{85}$Rb and $^{87}$Rb+$^{87}$Rb in Sec.~\ref{sec:numbers bound states}.

Eqs.~(\ref{eq:phaseintegral2}) and (\ref{eq:phaseintegral3}) enable us to mass-scale $\phi\phiE$ and $\phi\phiL$ for two isotopic pairs $\mA \equiv A_1,A_2$ and $\mA^\prime \equiv A_1^\prime,A_2^\prime$ ($A_i, A^\prime_i$ standing for atomic mass numbers) according to
\begin{equation}\label{eq:phiephil mass scaling}
^\mA \phi\phiE = \MR ~{^{\mA^\prime} \phi\phiE} \quad\mbox{and}\quad
^\mA \phi\phiL = \MR^{-1} ~ {^{\mA^\prime} \phi\phiL},
\end{equation}
where $\MR = \sqrt{\mu_\mA / \mu_{\mA^\prime}}$ with $\mu$ being a reduced mass. For these scaling equations contributions to $\phi(E,l)$ independent of $E$ and $l$ do not play a role. For the mass scaling of $\phi^0$, on the other hand, we have
\begin{equation}\label{eq:phi0parts}
\phi^0=n_b^{\prime} \pi + \phi^0_{mod(\pi)},
\end{equation}
with $n_b{^\prime}$ the number of zero-energy $s$-wave nodes up to the radius of interest ($r_0$), excluding the node at $r = 0$, and $\phi^0_{mod(\pi)}$ the modulo $\pi$ part of the total phase $\phi^0$. Each phase cycle $\pi$ corresponds to one additional radial node and thus an extra (vibrational) bound state in the potential.

Combining this equation with the mass scaling relation (\ref{eq:WKB Phi0 formula}) we find
\begin{equation}
^\mA \phi^0_{mod(\pi)} + {^{\mA}n_b^\prime} \pi - \frac{\pi}{4} =
\MR \left[^{\mA^\prime} \phi^0_{mod(\pi)} + {^{\mA^\prime}n_b^\prime} \pi - \frac{\pi}{4}\right],
\end{equation}
so that the scaled $\phi^0_{mod(\pi)}$ values of the two isotopic pairs are related according to
\begin{equation}\label{eq:mass scaling phi0}
^{\mA}\phi^0_{mod(\pi)} = \MR ~ ^{\mA^\prime}\phi^0_{mod(\pi)}
+ \left( 1 - \MR\right)
\frac {\pi}{4} - ~{^{\mA}n_b^\prime} \pi + \MR
~{^{\mA^\prime}n_b^\prime} \pi
\end{equation}
and its inverse, obtained by interchanging the isotopic atom pairs and substituting $1/\MR$ for $\MR$. The last term gives rise to a number of discrete values for the mass-scaled modulo $\pi$ phase of isotopic atom pair $\mA$, depending on $n_b^\prime$ for the other pair. The interval between these discrete values is $(1 - \MR)\pi$. This discretization can be exploited when extracting information from experimental data of multiple isotopic pairs and requiring the modulo $\pi$ phases for the pairs considered to be related according to Eq.~(\ref{eq:mass scaling phi0}). Clearly, this allows us to deduce $^{\mA^\prime}n_b^\prime$ and, by exchanging the roles of the isotope pairs, $^{\mA}n_b^\prime$. It should be emphasized that the (adiabatic) accumulated phase method thus offers a unique possibility to deduce numbers of bound states for potentials \textit{without knowing their short-range part up to $r_0$}. This approach has been applied in Ref.~\cite{vankempen} in the analysis of a set of experimental $^{85}$Rb and $^{87}$Rb bound state and cold collision data. In the present paper we build on that analysis, which we wish to describe and discuss in more detail. We come back to this in connection with column A of Table \ref{table1} that has been taken from \cite{vankempen}. In the same context we estimate the accuracy of the mass scaling for these isotopes.

\section{Accuracy of mass scaling}\label{sec:accuracy mass scaling}
A crucial issue for the possibility to combine the analysis of different isotope pairs is its expected accuracy. In that connection two types of corrections need discussion, corresponding to the adopted Born-Oppenheimer and WKB approximations.

\subsection*{Accuracy of mass scaling: adiabatic correction to BO}
The main correction to the Born-Oppenheimer approximation is the adiabatic or diagonal correction $V_{ad}$ to the interatomic potential~\cite{kutzelnigg}, given by
\begin{equation}\label{Vad}
V_{ad}(r) = < \psi_{el}(x;r)| -\frac{\hbar^2}{2\mu} \Delta_r|
   \psi_{el}(x;r) > \, \propto  \frac{1}{\mu},
\end{equation}
with $\psi_{el}$ the electronic wave function (x = electronic coordinates), depending parametrically on the nuclear coordinates. This leads to an adiabatic correction to the accumulated phase (\ref{eq:phaseintegral1}):
\begin{equation}
 \phi_{ad}(E,l) = - \frac{\mu}{\hbar^2} \int_{r_t}^{r_0} \frac{dr}{k(r)} V_{ad}(r)\label{eq:adiabatic phase change1}.
\end{equation}
To show its classical meaning we write it as a time integral over the collision in the classically allowed range within $r_0$ ($dt = dr/v(r[t])$):
\begin{equation}
\phi_{ad}(E,l) = - \frac{1}{\hbar} \int_{r_t}^{r_0} V_{ad}(r[t]) dt \\
\equiv - \frac{1}{\hbar} \tau_{col} < V_{ad} >_{cl},\label{eq:adiabatic phase change2}
\end{equation}
proportional to $1 / \sqrt \mu$. The last member of this equation indicates the proportionality to the collision time $\tau_{col}$ and to a classical expectation value in this range. In the following we estimate the isotopic spread $\Delta V_{ad}$ and thus the associated spread in accumulated phase parameters on the basis of experiment, on the basis of theory, and using a combination of both.

\textbf{1. Experimental evidence}

In 2000 a paper by Seto \emph{et al.}~\cite{seto} described a measurement of high-resolution $A \rightarrow X$ emission data for a mixture of the isotopic pairs $^{85}$Rb$_2$, $^{87}$Rb$_2$, and $^{85}$Rb$^{87}$Rb, covering in total 12148 transition frequencies. The data allowed a ground-breaking analysis of vibrational level spacings of the $X^1\Sigma^+_g$ electronic state up to $v = 113$ ($r$ up to 25$a_0$). Although the data set, with uncertainties $\pm$0.001 cm$^{-1}$, involved the above three isotopic pairs, the analysis turned out to lead to a common singlet potential without any sign of a Born-Oppenheimer breakdown. A similar analysis for the triplet case does not exist.

This result enables us to deduce an upper limit for the correction to a mass scaled singlet phase due to Born-Oppenheimer breakdown. To that end we consider the isotopic difference $\Delta \phi_{ad}(E,l)$ of the adiabatic phase correction and note that the above $\pm$0.001 cm$^{-1}$ uncertainties correspond to quantum mechanical expectation values of the isotopic difference $\Delta V_{ad}(r)$ over a large set of rovibrational states $v,l$ with probability densities covering together at least the whole range $[r_t,r_0]$. This justifies the conclusion that the isotopic difference $\Delta V_{ad}(r)$ is less than 0.001 cm$^{-1}$ in absolute value. For energies $E$ close to 0 and using Eq.~(\ref{eq:tau}), we thus find a correction due to the implicit isotopic dependence:
\begin{equation}\label{upperlimitexpS}
| \Delta\phi^0_S | \leq 0.001 \mbox{cm}^{-1} \phi^E_{S} = 0.33 \times 10^{-4}\pi.
\end{equation}
Here and in the following these estimates apply to the isotopic pairs $^{85,85}$Rb$_2$ - $^{87,87}$Rb and half these values to the pairs $^{85,85}$Rb$_2$ - $^{85,87}$Rb and $^{85,87}$Rb$_2$ - $^{87,87}$Rb. We have used the value of $\partial \phi_{S}/\partial{E} \equiv \phi^E_S$ from the analysis in Ref.~\cite{vankempen}. In the final result we have split off a factor $\pi$ representing the basic periodicity associated with the phases $\phi$. We expect a similar order of magnitude for the implicit isotopic correction in the triplet case.

\textbf{2. Theoretical evidence}

An order of magnitude estimate for both the singlet and triplet case can be based on the long-range expression for $V_{ad}$ proposed by Dalgarno and McCarroll  \cite{mccarroll}:
\begin{equation}\label{eq:dalgarno}
V_{ad} = - \frac{m_e}{4\mu} \left[V_{BO} + r \frac{dV_{BO}(r)}{dr} \right],
\end{equation}
with $m_e$ the electron mass and $V_{BO} \equiv V_{S/T}$ the Born-Oppenheimer potential for the atom pair. Assuming that (\ref{eq:dalgarno}) can be used for an order of magnitude estimate in the range $[r_t,r_0]$ \cite{jamieson}, we thus obtain
\begin{equation}\label{eq:adiabatic phase change}
\Delta\phi_{ad}(E,l) = -\frac{m_e}{4\hbar^2}\frac{\Delta\mu}{\mu} \int_{r_t}^{r_0} \frac{1}{k(r)} \left[V(r) + r \frac{dV(r)}{dr} \right] dr.
\end{equation}

With the singlet potential of Ref.~\cite{seto} and the ab-initio triplet potential from Ref.~\cite{kraussstevens} for $r < r_0$, both shifted 'vertically' and smoothly joined to dispersion $\pm$ exchange forms following from the parameters in the later Table \ref{table1} for $r > r_0$, we find
\begin{equation}\label{eq:upperlimitthST}
\Delta_V \phi^0_S = + 0.037 \times 10^{-5} \pi,\quad  \Delta_V \phi^0_T = -0.19 \times 10^{-5} \pi.
\end{equation}
We note that the smallness of the estimated singlet phase correction is due to the large negative contributions to the radial integral over the Dalgarno-McCarroll expression (\ref{eq:dalgarno}) at small $r$ values, which compensate the positive contributions at longer range to a considerable extent.

\textbf{3. Combined evidence}

To improve the above estimates on the basis of experiment and theory together, we note for the singlet case that the Dalgarno-McCarroll expression (\ref{eq:dalgarno}) is larger than the maximum adiabatic correction 0.001 cm$^{-1}$ in absolute value allowed by experiment~\cite{seto} in a range of atomic distances starting from the inner classical turning point $r_t$ = 5.9 $a_0$ until 7.7 $a_0$. We therefore use the experimental limit in the radial integral (\ref{eq:adiabatic phase change}) until a distance of 7.7 $a_0$ so that it fits continuously to the theoretical prediction in the further interval up to the final radius $r_0$ = 16 $a_0$. For the triplet situation $r_t$ is much larger (about 9.5 $a_0$). In that case the Dalgarno-McCarroll expression is smaller in absolute value than 0.001 cm$^{-1}$ over the whole interval $[r_t,r_0]$. Substituting that in the radial integral, we find our triplet result. In total we find
\begin{equation}\label{eq:upperlimitexpthST}
|\Delta \phi^0_S| = 0.61 \times 10^{-5} \pi,\quad  |\Delta \phi^0_T| = 0.19 \times 10^{-5} \pi.
\end{equation}

\subsection*{Accuracy of mass scaling: corrections to WKB}
The order of magnitude of this correction is easily estimated by comparing the mass scaled $^{85}$Rb phase parameters to those obtained by numerical integration of the singlet and triplet radial Schr\"{o}dinger equations up to $r = 16a_0$ for the above-mentioned singlet and triplet potentials with the reduced masses involved. The deviations of the mass scaled phases are
\begin{equation}\label{eq:correction wkb}
|\Delta \phi^0_S| = |\Delta \phi^0_T| = 2 \times 10^{-5} \pi.
\end{equation}

\begin{table}
\begin{center}
\caption{Interaction parameters (au) derived from combined $^{85}$Rb and $^{87}$Rb experiments (column A) including error bars, mainly due to 10\% uncertainty in $C_{10}$; column B: fractional changes due to phase corrections; column C: percentages of variation of same quantities over range [10.85,16] $a_0$ of $r_0$ values according to adiabatic accumulated phase method; column D: same for conventional method.} \label{table1}
\begin{tabular}[t]{ccccc} \hline \hline
 Quantity & A & B(\%) & C(\%) & D(\%) \\
     \hline
$C_6/10^3$        & 4.703(9) & 0.001 & 0.04 & 0.1    \\
$C_8/10^5$        & 5.79(49) & 0.002 & 0.2 & 0.6  \\
$C_{10}/10^7$     & 7.665(Ref.~\cite{marinescu})  &   &   \\
$J.10^2$          & 0.45(6) & 3 & 1 & 2   \\
$a_T(^{87}$Rb)    & +98.98(4)&  0.0004 & 0.001 & 0.02   \\
$a_S(^{87}$Rb)    & +90.4(2) & 0.02 & 0.09 & 0.2 \\
$a_T(^{85}$Rb)    & -388(3) & 0.06 & 0.2 & 0.3 \\
$a_S(^{85}$Rb)    & +2795$^{+420}_{-290}$  & 0.5 & 3 & 7 \\
$v_{DT}$(mod 1),$n_{bT}(^{87}$Rb) & 0.4215(3),\ 41 & 0.001 & 0.03 & 0.04  \\
$v_{DS}$(mod 1),$n_{bS}(^{87}$Rb) & 0.455(1),\ 125 & 0.02 & 0.07 & 0.10 \\
$v_{DT}$(mod 1),$n_{bT}(^{85}$Rb) & 0.9471(2),\ 40 & 0.002 & 0.008 & 0.02\\
$v_{DS}$(mod 1),$n_{bS}(^{85}$Rb) & 0.009(1),\ 124 & 0.5 & 3 & 7  \\
\hline \hline
\end{tabular}
\end{center}
\end{table}

\subsection*{Comparison of phase corrections to error bars from previous analysis}

To illustrate the smallness of the above estimated phase corrections, we compare them with the error bars obtained in our previous brief description of the adiabatic accumulated phase method in Ref.~\cite{vankempen}. In that Letter a combined analysis of $^{85}$Rb and $^{87}$Rb experimental data led to values for interaction and scattering properties of Rb atoms with an unprecedented accuracy. In column A of Table \ref{table1} we recapitulate the dispersion coefficients $C_6$, $C_8$, the strength parameter J of the exchange interaction, and the set of pure singlet and triplet scattering lengths + associated fractional vibrational quantum numbers at dissociation $v_D$, together with their error bars. Column B gives for comparison the maximum fractional changes (in \%) of the same quantities that result from the combination of the two types of phase corrections above. We conclude that the latter are small compared to the error bars resulting from the analysis in Ref.~\cite{vankempen} and indicated in column A. The latter are mainly due to the 10 \% error assumed for the theoretical $C_{10}$ value taken from Ref.~\cite{marinescu}. The largest of the fractional phase corrections is that for $J$. We note that that is not unexpected taking into account that this concerns the coefficient of a radially exponential term, which is extremely sensitive to the damping coefficient in the exponential. This also explains the relatively large error bar in column A.

The beautiful agreement with experiment, achieved in the analysis of Ref.~\cite{vankempen}, is a convincing further indication that the mass scaling procedure is an excellent approximation. For instance, the values of $C_6$ and $C_8$ agree with values $C_6 = 4.691(23) \times 10^3$ \cite{derevianko} and $C_8 = 5.77(8) \times 10^5$ \cite{derevianko1}, calculated by Derevianko and co-workers, while $J$ agrees with the most recent calculated value $J = 0.384 \times 10^2$ published by Hadinger and Hadinger \cite{hadinger}.

We can also conclude that there is considerable room for an extension of the mass scaling procedure to applications of the adiabatic accumulated phase method to isotopic pairs of lighter elements than the Rb isotopes studied here, despite the expected larger phase corrections due to Born-Oppenheimer and WKB breakdown.

\section{Comparison of adiabatic to conventional accumulated phase method and dependence on $r_0$}\label{sec:choice r0}

To illustrate the advantages of our adiabatic accumulated phase
method, we compare a calculation including the adiabatic spin
mixing at $r_0$ to one without, i.e., the conventional approach. In
both cases we consider the optimization of the potential parameters
given a set of $^{85}$Rb and $^{87}$Rb experimental data according to the analysis in Ref.~\cite{vankempen}. It turns out that the optimized values of the quantities in Table \ref{table1} are highly independent of the choice of $r_0$. To demonstrate that we have added the percentages of variation over the $r_0$ interval [10.85,16.0] in column C of the table. In this case too the exchange strength parameter $J$ is an exception, with a variation of 1\%. This can be explained as indicated above in connection with columns A and B of Table \ref{table1}. In column D we have added for comparison the significantly larger percentages of variation of the same quantities according to the conventional accumulated phase method.

\begin{figure}
\begin{center}
\includegraphics[width=0.45\textwidth]{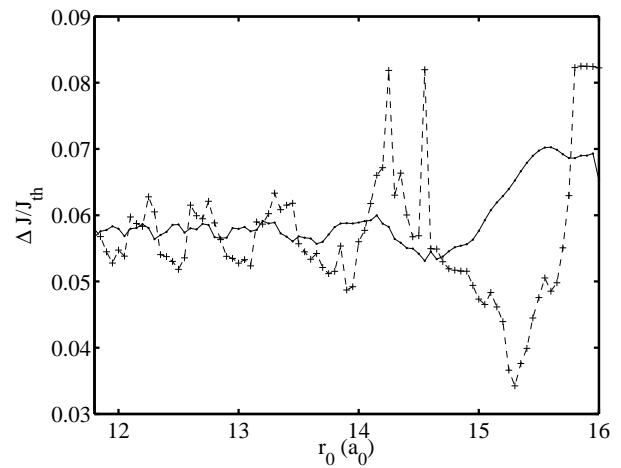}
\end{center} \caption{Fractional correction to the literature value $J_{\mr{th}}$ for the strength parameter of the exchange energy versus $r_0$~\cite{hadinger}. The dashed line connects points calculated with the traditional accumulated phase method, the solid curve to those using the adiabatic accumulated phase method.} \label{graph:ParvsR0}
\end{figure}

In Fig.~\ref{graph:ParvsR0} we select by far the most sensitive parameter, $J$, to show its $r_0$ dependence. Each of the curves shows the behavior of the fractional deviation $\Delta J/J_{\mr{th}}$ of the optimal $J$ value from the value $J_{\mr{th}}$ published by Hadinger \emph{et al}.~\cite{hadinger}. The + signs connected by the dashed curve show the result of a calculation along conventional lines. Each point indicated on the curve represents the outcome of a separate $\chi^2$ optimization. Switching on the spin mixing adiabatically at $r_0$ gives rise to the solid line. Clearly, the oscillation is strongly reduced. The remaining oscillation is mainly due to the WKB correction and the non-adiabaticity of switching on the coupling due to $V^{hf-}$. Even shifting $r_0$ to 16$a_0$ keeps the oscillation amplitude in $J$ to below the 1\% level.

Fig.~\ref{graph:ParvsR0} suggests that one might just as well select a smaller value for $r_0$ near 12 $a_0$ to avoid the $V^{hf-}$ coupling issue altogether. If we would have done that from the beginning, however, we would have missed a key message from our study: the fact that the final results are highly independent of the central potentials within an interatomic distance of 16$a_0$. This applies in particular to the exchange potential $V\exchange$ for which the Smirnov-Chibisov radial dependence (\ref{eq:Vexch}) is an asymptotic expression. The same applies to the asymptotic expression (\ref{eq:Vdisp}) for the dispersion potential.

\section{Determining numbers of singlet and triplet bound states for $^{85}$Rb + $^{85}$Rb and $^{87}$Rb + $^{87}$Rb systems}\label{sec:numbers bound states}

Here we come back to the relation between the mass-scaled modulo $\pi$ accumulated phases for different isotopic versions of a general atom-atom system discussed in Sec.~\ref{sec:mass scaling}, in particular Eq.~(\ref{eq:mass scaling phi0}). This relation and its inverse contain the (unknown) numbers of nodes $n_b^\prime$ of the zero-energy radial wave function contained in the potential from the inner turning point up to $r_0$ for the two interrelated atom pairs $\mA$ and $\mA^\prime$. As pointed out above, this enables us to deduce the total numbers of bound singlet and triplet states from available experimental data.

To illustrate our method via the example of the $^{85}$Rb + $^{85}$Rb and $^{87}$Rb + $^{87}$Rb systems, it is helpful to recapitulate some aspects of the analysis that led to column A of Table \ref{table1} in Ref.\cite{vankempen}, reproduced in the above Table \ref{table1}. The experimental material analyzed consisted of data on cold collisions and on bound states exceptionally close to the continuum, partly for $^{85}$Rb and partly for $^{87}$Rb. The six parameters varied in a $\chi^2$ analysis were $^{87}\phi_T^0$, $^{87}\phi_T^E$, $^{87}\phi_T^l$, $C_6$, $C_8$, and $J$, with $C_{10}$ held fixed at the theoretical value from Marinescu \emph{et al}.~\cite{marinescu}.

\begin{figure}
\begin{center}
\includegraphics[width=0.45\textwidth]{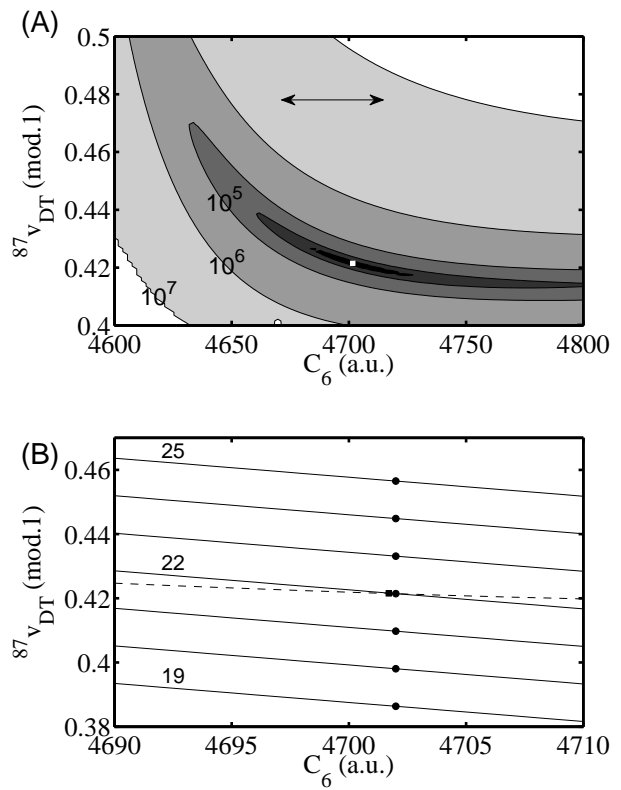}
\end{center} \caption{(A) Contour plot of the reduced $\chi^2$ as a function of $C_6$ and $^{87}v_{DT}$(mod 1) for $^{87}$Rb data only. The dispersion coefficient $C_6$ is expected to be within the interval indicated by the horizontal double arrow \cite{derevianko}. (B) The dashed line indicates the bottom of de 'deepest trench' in the $\chi^2$ surface of part (A), with the overall minimum indicated by the square. The solid lines indicate equivalent trenches in the $\chi^2$ surface for $^{85}$Rb data only, making use of mass scaling to deduce $^{85}\phi_T^0$(mod $\pi$) from $^{87}v_{DT}$(mod 1). The position of the minimum $\chi^2$ value along each trench is indicated by a solid circle. Each of the solid lines corresponds to a specific number $^{87}n_b^{\prime}$ of nodes assumed to be contained in the $^{87}$Rb triplet potential up to $r_0$: $^{87}n_b^{\prime}$ = 19,....,25. Both graphs A and B have been generated by varying $^{87}v_{DT}$(mod 1) and $C_6$, keeping the remaining fit parameters $^{87}\phi^E$, $^{87}\phi^l$, $J$, and $C_8$ of the $\chi^2$ analysis in Ref.~\cite{vankempen} fixed at their optimal values.} \label{graph:Chi2_Vdt87_C6}
\end{figure}

Instead, in part A of Fig.~\ref{graph:Chi2_Vdt87_C6} we present a contour plot of the reduced $\chi^2$ value for the subset of $^{87}$Rb data as a function of $C_6$ and $^{87}v_{DT}$(mod 1) [equivalent to $^{87}\phi_T^0$(mod $\pi$)], keeping  the remaining parameters $^{87}\phi_T^E$, $^{87}\phi_T^l$, $C_8$, and $J$ fixed at their values according to column A. The dispersion coefficient $C_6$ is expected to be within the interval indicated by the horizontal double arrow \cite{derevianko}. The absolute $\chi^2$ minimum is the white point indicated within the $< 10^5$ area. Part B is based on the $^{85}$Rb data only, again keeping $^{87}\phi_T^E$, $^{87}\phi_T^l$, $C_8$, and $J$ fixed and calculating the three $^{85}$Rb phase parameters by mass scaling. The dashed line shows the bottom of the 'deepest trench' in the $\chi^2$ surface of part A, with the overall minimum indicated by the square. The solid lines indicate equivalent trenches in the $\chi^2$ surface for the $^{85}$Rb data. Note that the interval between the discrete $^{85}$Rb phase values in Eq.~(\ref{eq:mass scaling phi0}) is approximately $(1-\MR)\pi \approx 0.012\pi \approx 0.036$, corresponding to intervals $(1-\MR) \approx 0.012$ for $^{87}v_{DT}$ between the solid lines. The position of the minimum $\chi^2$ value along each trench is indicated by a solid circle. Each of the solid lines corresponds to a specific number $^{87}n_b^{\prime}$ of nodes assumed to be contained in the $^{87}$Rb triplet potential up to $r_0$ = 16$a_0$: $^{87}n_b^{\prime}$ = 19,....,25. Clearly, 22 is the preferred value for $^{87}n_b^{\prime}$. It turns out that the lower mass in the $^{85}$Rb case does not decrease this value:  $^{85}n_b^{\prime}$ = 22. With the optimalized dispersion and exchange parameters of Table \ref{table1} we can continue the solution of the zero energy triplet radial Schr\"{o}dinger equations for the two atom pairs to find the total numbers of bound states. For the case of $^{87}$Rb we thus find the total number of triplet bound states $n_{bT}(^{87}$Rb$_2$) = 41. Similarly for $^{85}$Rb:  $n_{bT}(^{85}$Rb$_2$) = 40. With the singlet potential of Seto \emph{et al.}, suitably shifted vertically and also completed with a long range part the result is $n_{bS}(^{87}$Rb$_2$) = 125, $n_{bS}(^{85}$Rb$_2$) = 124.

\section{Summary and outlook}\label{sec:summary}

We have presented a theoretical method that enables one to describe and predict the interaction and scattering properties of (ultracold) atoms. It allow us, for instance, to predict the $^{87}$Rb spinor condensate to be ferromagnetic \cite{vankempen}, a prediction for which the relevant scattering lengths have to be calculated with a precision better than 1\%. It is also unparalleled in comprehensiveness: it allows the prediction of a large and varied set of experimental quantities for all pairs of like and unlike atoms.

Its original version, the accumulated phase method, was designed to predict essential properties like scattering lengths and Feshbach resonances, enabling the realization of Bose-Einstein condensates and Fermi degenerate gases of alkali atoms, for which the short-range interaction was insufficiently known to calculate these properties directly. The method consisted of replacing the short-range interaction with a boundary condition on the two-atom wave function at an interatomic distance $r = r_0$, deducing the boundary condition from available experimental data, and predicting all other relevant data. The new, \textit{adiabatic}, version of the method, described in the present paper, has been presented briefly in a previous Letter\cite{vankempen}. Whereas the original method neglected the hyperfine coupling between singlet and triplet states for $r < r_0$ and included this coupling together with asymptotic dispersion + exchange expressions for $r > r_0$, the new approach takes the adiabatic singlet-triplet mixing by $V_{hf}$ into account at the separation radius $r_0$ and therefore effectively also at smaller $r$, neglecting the (second order) changes of the radial waves. This makes it possible to shift $r_0$ to larger interatomic distances, thus allowing for more reliable asymptotic potential terms in the range $r > r_0$.

We have described how the second-order spin-orbit interaction can be included, as well as a mass scaling approach to relate the accumulated phases for different isotopic versions of atom pairs. The accuracy of the mass scaling has been discussed, taking into account both Born-Oppenheimer and WKB breakdown. Estimates have been given for the Rb isotopes, pointing to a high accuracy. Using the Rb isotopes for illustration, the adiabatic and conventional accumulated phase methods were compared, and the $r_0$ dependence of their optimized interaction parameters was studied. Finally, we have explained how the total numbers of bound singlet and triplet two-atom states follow from a combined analysis of different isotopic versions of atom pairs, without knowing the short range interatomic interaction.

We believe that the new adiabatic accumulated phase method has great potential to support further studies of cold atom systems, especially in the rapidly growing field of pairs of unlike atoms, to which the method can readily be extended \cite{inguscio}. The set of phase parameters that it makes use of can be systematically extended when larger energy or angular momentum ranges come into play experimentally, contrary to other choices used for the adjustment of the short-range part of model potentials\cite{tiesinga1,julienne,gao}.

\section{Acknowledgements}
This work was supported by the Netherlands Organization for Scientific Research (NWO). E.G.M.~van Kempen acknowledges support from the Stichting FOM, which is financially supported by NWO.


\begin{thebibliography}{100}
\bibitem{remark} Fig.~\ref{graph:vsort44} was originally included in Ref.~\cite{kokkelmans1}. Due to remarks by one of the referees we were forced to leave it out.
\end{thebibliography}
\end{document}